\documentclass[
    ,final
  ,numberedheadings 
  ,cmfonts          
  ,float
  ,subfigure
  ]
  {aipproc}

\usepackage{amsmath}
\usepackage{subfigure}
\layoutstyle{6x9}

\begin{document}

%\preprint{}

\title{Dispersive approach  in Sudakov resummation}


\author{Georges Grunberg}{
address={Centre de Physique Th\'eorique, \'Ecole polytechnique, CNRS,  \\
91128 Palaiseau, France\\
E-mail: grunberg@cpht.polytechnique.fr}}

\begin{abstract}
The dispersive approach to power corrections is given a precise implementation, valid beyond single gluon exchange, in the framework of Sudakov resummation for deep inelastic scattering and the Drell-Yan process. It is shown that the assumption of infrared finite Sudakov effective couplings implies the universality of the corresponding infrared fixed points. This property is closely tied to the universality of the virtual contributions to space-like and time-like processes, encapsulated in the second logarithmic derivative of the quark form factor.
\end{abstract}

\maketitle

\noindent The infrared (IR) finite coupling (``dispersive'') approach to power corrections \cite{DMW} provides an
attractive framework where the issue of universality can be meaningfully raised. This approach however seems to be
tied in an
 essential way to the single gluon exchange approximation. In this talk   I show that it can actually find a precise implementation in the framework of Sudakov
resummation, and that its validity extends beyond single gluon exchange.

\noindent Consider first the scaling violation in deep inelastic scattering (DIS) in Mellin space at large $N$. One can show
\cite{Gru-talk} that Sudakov resummation takes in this case the very simple form 

\begin{equation}{d\ln F_2(Q^2,N)\over d\ln Q^2}=4 C_F\int_{0}^{Q^2}{dk^2\over k^2} G(N k^2/Q^2)
A_{{\cal S}}(k^2)
+4 C_F H(Q^2)+{\cal O}(1/N)  
\label{eq:scale-viol},\end{equation}
where   the ``Sudakov effective coupling'' $A_{{\cal S}}(k^2)$, as well as $H(Q^2)$,  are  given as  power series in
$ \alpha_s$ with $N$-independent coefficients. In the standard resummation framework one has
$A_{{\cal S}}(k^2)=A_{{\cal S}}^{stan}(k^2)$ with
$4 C_F A_{{\cal S}}^{stan}(k^2)=A(\alpha_s(k^2))+dB(\alpha_s(k^2))/d\ln
k^2$
 where
$A$ (the universal  ``cusp'' anomalous dimension) and $B$ are  the standard Sudakov anomalous dimensions
relevant to DIS, and
$G(N k^2/Q^2)=G_{stan}(N k^2/Q^2)\equiv \exp(-N k^2/Q^2)-1$.
 It was further  observed in \cite{Gru-talk} that the separation
between the constant terms contained in the Sudakov integral on the right hand side of eq.(\ref{eq:scale-viol})
 and the ``leftover'' constant terms contained in $H(Q^2)$ is arbitrary,
yielding a variety of Sudakov resummation procedures, different choices leading to a different ``Sudakov distribution
function''
$G(N k^2/Q^2)$ and effective coupling
$A_{{\cal S}}(k^2)$, as well as to a different function
$H(Q^2)$.   This  freedom of selecting the constant terms actually  disappears by
taking one more derivative, namely

\begin{equation}{d^2\ln F_2(Q^2,N)\over (d\ln Q^2)^2}=4 C_F\int_{0}^{\infty}{dk^2\over k^2}
\dot{G}({N k^2\over Q^2}) A_{{\cal S}}(k^2)
+4 C_F[{dH\over d\ln Q^2}-A_{{\cal S}}(Q^2)]+{\cal
O}({1\over N})\label{eq:d-scale-viol},\end{equation}
where $\dot{G}=- dG/ d\ln k^2$. The point is that the Sudakov integral ${\cal
S}(Q^2,N)$ on the right hand side of
eq.(\ref{eq:d-scale-viol}) being UV convergent, all the large $N$ logarithmic terms are now determined by the
${\cal O}(N^0)$ terms contained in the integral, which therefore cannot be fixed arbitrarily anymore. Thus ${\cal S}(Q^2,N)$ is uniquely determined.
This observation implies in turn that  the combination
$dH/ d\ln Q^2-A_{{\cal S}}(Q^2)$, which represents the ``leftover'' constant terms not included in ${\cal
S}(Q^2,N)$, is also  fixed. In fact, I conjecture that it is related to the space-like on-shell
electromagnetic quark form factor
\cite{MVV} ${\cal F}_q(Q^2)$ by
  
\begin{equation}4 C_F\left({dH\over d\ln Q^2}-A_{{\cal S}}(Q^2)\right)={d^2\ln ({\cal F}_q(Q^2))^2 \over
(d\ln Q^2)^2}\label{eq:form-factor}.\end{equation}
Eq.(\ref{eq:form-factor}) has been checked \cite{Gru-Friot} to  ${\cal O}(\alpha_s^4)$. 
For the short distance Drell-Yan cross
section, the analogue of  (\ref{eq:form-factor}) is 
$4 C_F\left({dH_{DY}\over d\ln Q^2}-A_{{\cal S},DY}(Q^2)\right)={d^2\ln \vert{\cal F}_q(-Q^2)\vert^2
\over (d\ln Q^2)^2}\label{eq:form-factor-DY}$
where ${\cal F}_q(-Q^2)$ is the time-like quark form factor.

\noindent However, although 
${\cal S}(Q^2,N)$  is  uniquely determined, the Sudakov distribution function and effective
coupling are still {\em not}.  We  deal with an infinite variety of different representations of ${\cal S}(Q^2,N)$, all equivalent for
the purpose of resumming Sudakov logarithms.
A prescription to  single out the correct representation  relevant for the issue of power corrections in the IR
finite coupling approach is needed. In absence of
the appropriate criterion,  predictions such as existence of an ${\cal O}(1/Q)$ linear power correction
\cite{KS} in Drell-Yan, or logarithmically-enhanced power corrections \cite{Gru-talk}, which follow from
particular choices of
$G$, cannot be a priori dismissed.  
In the last paper of  \cite{Gru-talk}, it was suggested to select the correct $G$ by requiring the corresponding $A_{\cal S}$ to be identical at large $N_f$ to the {\em Minkowskian} coupling defined as
the time-like (integrated) discontinuity of the {\em Euclidean} one-loop coupling (the so-called
``V-scheme'' coupling) associated to the  dressed gluon propagator:
$A_{{\cal
S},\infty}^{Mink}(k^2)={1\over\beta_0}\left[{1\over
2}-{1\over\pi}\arctan(t/\pi)\right]$
with $t=\ln (k^2/\Lambda^2_V)$ (where $\Lambda_V$ is the V-scheme scale parameter). As shown in \cite{Gru-talk}, this ansatz
fixes the corresponding ``Minkowskian''  Sudakov distribution function (which one could also call ``characteristic
function'' following
\cite{DMW}) to be given in the DIS case by
$G_{Mink}(N k^2/Q^2)=
\ddot{{\cal G}}(\epsilon)$, with
$\epsilon=N k^2/Q^2$. The function $\ddot{{\cal G}}(\epsilon)$ is
obtained from the finite $N$ characteristic function \cite{DMW} ${\cal F}(\lambda^2/Q^2,N)$ (where
$\lambda$ is the ``gluon mass'')
 by defining
${\cal G}(y,N)\equiv{\cal F}(\lambda^2/Q^2,N)$ with $y\equiv N \lambda^2/Q^2$, and taking the
$N\rightarrow\infty$ limit at {\em fixed} $y$:
$\ddot{{\cal G}}(y,N)\rightarrow \ddot{{\cal G}}(y,\infty)\equiv
\ddot{{\cal G}}(y)$ (where $\dot{{\cal G}}=- d{\cal G}/ d\ln Q^2$).
In the Drell-Yan case, the same requirement yields 
instead $G_{Mink}^{DY}(N k/Q)=\ddot{{\cal G}}_{DY}(\epsilon_{DY}^2)$, with $\epsilon_{DY}=N k/Q$
. Similarly,
$\ddot{{\cal G}}_{DY}(y_{DY})$ is obtained by taking the large $N$ limit at fixed $y_{DY}= N^2 \lambda^2/Q^2$ of
the finite $N$ characteristic function
\cite{DMW} ${\cal F}_{DY}(\lambda^2/Q^2,N)$.   The same Sudakov distribution function $G_{Mink}^{DY}(N k/Q)$  also
follows from the resummation formalism (not tied to the single gluon approximation) of
\cite{Vogelsang}, which  therefore uses an implicitly Minkowskian framework in the above sense.
 
\noindent
Since they are $N_f$-independent, the {\em same}  Minkowskian Sudakov distribution functions, now fixed
through the large
$N_f$ identification of the Sudakov effective couplings, can then be used to determine the corresponding  effective couplings at {\em finite}
$N_f$ in the usual way, requiring  the large $N$ logarithmic terms on the left hand side of
eq.(\ref{eq:scale-viol})  (with  $G=G_{Mink}$)
 to be correctly reproduced order by order in
perturbation theory. This proposal is  equivalent to generalize the basic equation of the dispersive approach \cite{DMW}  to {\em all orders} in $\alpha_s$ in the large $N$ limit to the statement that ${d\ln F_2(Q^2,N)\over d\ln Q^2}=4 C_F\int_{0}^{\infty}{d\lambda^2\over \lambda^2}
\ddot{{\cal F}}({ \lambda^2\over Q^2},N) A_{{\cal S}}^{Mink}(\lambda^2)
+4C_F \Delta H_{Mink}(Q^2)+{\cal
O}({1\over N})$ (a ``left-over'' $\Delta H_{Mink}(Q^2)$ contribution is still expected at finite $N_f$). 
It is  natural to keep referring to the resulting $A_{{\cal S}}^{Mink}(\lambda^2)$  coupling  
  as Minkowskian   even at {\em finite} $N_f$, where
 identification to a dressed gluon propagator is no longer possible. In the Minkowskian formalism, only {\em non-analytic} terms in the small ``gluon mass'' expansion of the
characteristic function do contribute \cite{DMW} to the IR power corrections through their discontinuities, and power corrections are given by low-energy moments of the corresponding \cite{Gru-talk} Euclidean coupling $A_{{\cal S}}^{Eucl}(k^2)$. Thus consistency with IR renormalons expectations is guaranteed.

\noindent\underline{Universality issues}:
the IR finite coupling approach allows some statement on the universality of power corrections to various
processes. Indeed,
at large
$N_f$ there is universality to all orders in $\alpha_s$  between $A_{{\cal S}}^{Eucl}(k^2)$
and
$A_{{\cal S},DY}^{Eucl}(k^2)$, which in the present framework are both prescribed to be equal to the  one-loop
V-scheme coupling in this limit. At finite $N_f$ however it easy to check that  universality in the ultraviolet
region  holds only up to next to leading order in $\alpha_s$, where the DIS and Drell-Yan Euclidean  Sudakov
effective couplings actually coincide
 (up to a 
$1/4C_F$ factor) with the ``cusp'' anomalous dimension $A(k^2)$, but is lost beyond that order.

\noindent On the other hand, an interesting universality property holds in the IR region at finite $N_f$,
{\em assuming} the Sudakov effective couplings reach  non-trivial IR fixed points  at zero momentum. Indeed, eq.(\ref{eq:form-factor}) shows that for a given selection of ``leftover'' constant
terms (contained in the renormalization group invariant
 function
$H(Q^2)$), the corresponding Sudakov effective coupling $A_{{\cal S}}(Q^2)$ differs from the {\em universal}
quantity (for space-like processes)
$A_{{\cal S}}^{all}(Q^2)\equiv -{1\over 4 C_F}{d^2\ln ({\cal F}_q(Q^2))^2 \over
(d\ln Q^2)^2}$
only by the total   derivative   $dH/d\ln Q^2$.
The latter is again expected to vanish at zero momentum if one assumes $H(Q^2)$ also reaches a
non-trivial IR fixed point
$H(0)$. A similar argument applies to 
$A_{{\cal S},DY}(Q^2)$, with $A_{{\cal S}}^{all}(Q^2)$ replaced by its
time-like counterpart $Re[A_{{\cal S}}^{all}(-Q^2)]=-{1\over 4 C_F}{d^2\ln \vert{\cal F}_q(-Q^2)\vert^2 \over
(d\ln Q^2)^2}$. Thus we expect, for {\em any} resummation procedure
(assuming the IR fixed points exist) $A_{{\cal S}}(0)= A_{{\cal
S},DY}(0)=A_{{\cal S}}^{all}(0)$.

\bibliography{apssamp}% Produces the bibliography via BibTeX.

\begin{thebibliography}{9}

\bibitem{DMW} Yu.L. Dokshitzer, G. Marchesini and B.R. Webber,  
{\em Nucl. Phys.} {\bf B469} 
(1996) 93, and references therein; Yu.L. Dokshitzer and B.R. Webber, Phys.Lett. {\bf B404} (1997) 321.




\bibitem{Gru-talk} G. Grunberg, hep-ph/0601140; {\em Phys.Rev.} {\bf D73} (2006) 091901(R); hep-ph/0609309.







\bibitem{MVV} S. Moch, J.A.M. Vermaseren and A. Vogt, {\em JHEP} {\bf 0508} (2005) 049, and references therein.



\bibitem{Gru-Friot} S. Friot and G. Grunberg, in preparation.





         


\bibitem{KS} G.P. Korchemsky and G. Sterman, {\em
Nucl. Phys.} {\bf B437} (1995) 415.

\bibitem{Vogelsang} E. Laenen, G. Sterman and W. Vogelsang,
hep-ph/0010183; {\em Phys. Rev.} {\bf D63} (2001) 114018.
































\end{thebibliography}

\end{document}